\documentclass[12pt]{article}

\usepackage{amssymb}
\usepackage{amsmath}
\usepackage{amscd}
\usepackage{latexsym}
\usepackage{graphicx}

\usepackage{cite}
\newcommand{\be}{\begin{equation}}
\newcommand{\ee}{\end{equation}}
\newcommand{\Dlt}{\Delta}
\newcommand{\dlt}{\delta}
\newcommand{\prt}{\partial}
\newcommand{\br}{{\bf r}}
\newcommand{\bk}{{\bf k}}

\newcommand{\bt}{\beta}

\newcommand{\ep}{\varepsilon}
\newcommand{\al}{\alpha}
\newcommand{\ra}{\rightarrow}
\newcommand{\sgm}{\sigma}

\newcommand{\gm}{\gamma}
\newcommand{\om}{\omega}
\newcommand{\Om}{\Omega}

\newcommand{\dgr}{\dagger}

\newcommand{\rgl}{\rangle}
\newcommand{\lgl}{\langle}

\begin{document}

\begin{center}

{\Large{\bf Stability of normal quantum-fluid mixtures} \\ [5mm]

V.I. Yukalov$^{1,2}$ }  \\ [3mm]

{\it
$^1$Bogolubov Laboratory of Theoretical Physics, \\
Joint Institute for Nuclear Research, Dubna 141980, Russia \\ [2mm]

$^2$Instituto de Fisica de S\~ao Carlos, Universidade de S\~ao Paulo, \\
CP 369, S\~ao Carlos 13560-970, S\~ao Paulo, Brazil} \\ [2mm] 

{\bf E-mail}: {\it yukalov@theor.jinr.ru}

\end{center}

\vskip 1cm

\begin{abstract}

Mixtures of quantum fluids, that is gases or liquids, are considered with the 
emphasis on the conditions characterizing the stability of the mixtures. The 
mixtures, that can be formed by cold atoms or molecules, are assumed to be 
quantum requiring the description using quantum techniques, but not so cold 
that to exhibit superfluidity or superconductivity. Reviewing the stability 
conditions of such normal quantum systems is important for the comparison of 
these conditions with the stability conditions of, e.g., Bose-condensed mixtures. 
The behavior of observable quantities under the stratification of quantum mixtures 
is discussed.   

\end{abstract}


\newpage

\section{Introduction}

Quantum systems composed of atoms or molecules are nowadays intensively studied
due to the developed techniques of cooling and trapping \cite{Letokhov_1}.
The major interest has been paid to the study of Bose systems in the condensed 
state, which several books \cite{Lieb_2,Pethick_3,Ueda_4} and review articles 
\cite{Courteille_5,Andersen_6,Yukalov_7,Bongs_8,Yukalov_9,Posazhennikova_10,Morsch_11,
Yukalov_12,Moseley_13,Bloch_14,Proukakis_15,Yurovsky_16} are devoted to. Of particular 
interest are the mixtures of quantum superfluids, Bose-Bose mixtures \cite{Castilho_17},   
Bose-Fermi mixtures \cite{Saarela_18}, and Fermi-Fermi superfluid mixtures 
\cite{Adhikari_19}. 

One of the main points in considering quantum mixtures is the problem of their 
stability with respect to their stratification into separate components occupying 
different spatial locations. This type of stability, for brevity, is called 
{\it mixture stability}. The mixture stability of superfluid Bose-Bose mixtures 
is usually analyzed in the coherent approximation \cite{Yukalov_20,Abad_21}
or Bogolubov approximation \cite{Nepomnyashchii_22}. The Hugenholtz-Pines relation
\cite{Hugenholtz_23} for superfluid systems can be generalized for superfluid 
mixtures \cite{Nepomnyashchii_24,Watabe_25}. The stability of superfluid mixture 
in the Hartree-Fock-Bogolubov approximation has been recently analyzed 
\cite{Rakhimov_26}.      

The mixture stability for normal (not superfluid) quantum fluids has been considered 
to a much lesser extent. The aim of the present paper is to compensate this 
deficiency and to survey what is known on the stability conditions of normal quantum 
mixtures. 

The principal difference in the study of mixture stability for superfluid and normal 
fluids is in the following. The single-particle spectrum of a superfluid, where the 
gauge symmetry is broken, coincides with the spectrum of collective excitations
\cite{Gavoret_27,Watabe_28}. Therefore, to investigate the dynamic stability of a 
superfluid spectrum it is sufficient to consider the poles of the single-particle 
Green function. Contrary to this, for a normal quantum system, where the gauge symmetry 
is preserved, it is necessary to study the poles of the two-particle Green function or 
of the response function (dynamic susceptibility). For a normal system, the 
single-particle Green function does not define the spectrum of collective excitations. 

It is interesting and important to compare the conditions of mixture stability for
superfluid and normal quantum systems. Also, it is useful to compare dynamic and 
thermodynamic stability \cite{Kurten_29,Yukalov_30}. Dynamic stability requires that 
the spectrum of collective excitations be positive for finite momenta. And 
thermodynamic stability studies conditions where the thermodynamic potential of a
mixture is minimal. 

We consider the systems composed of neutral atoms or molecules, whose density 
fluctuations enjoy sound modes. The particles can be bosons or fermions, provided
that sound modes are well defined for them. The instability of a mixture with respect
to the spatial separation of the components can be caused by the interspecies 
interactions. The instability can arise when the repulsive interspecies interactions 
become sufficiently strong. The derivation of conditions characterizing the mixture 
stability of normal quantum-fluid mixtures is the main topic of the present survey. 

Throughout the paper, the system of units is employed where the Planck and Boltzmann
constants are set to one.

\section{Quantum-fluid mixture}

Let us start with the analysis of dynamic stability of a quantum-fluid mixture. 
We consider particles without internal degrees of freedom, such as spin. In the case
of particles with spin, this could imply that their spins are frozen in one direction 
and thus are switched off the particle dynamics.  

The grand Hamiltonian of the mixture of several components, enumerated by the
index $i$, has the form
$$
H = \sum_i \int \psi_i^\dgr(\br) \; \left( -\; \frac{\nabla^2}{2m} + U_i - 
\mu_i \right) \; \psi_i(\br) \; d\br \; +
$$
\be
\label{1}
 +\;
\frac{1}{2} \sum_{ij} \psi_i^\dgr(\br) \; \psi_j^\dgr(\br') \; 
\Phi_{ij}(\br-\br') \; \psi_j(\br') \; \psi_i(\br) \; d\br d\br' \; ,
\ee
where $\psi_i({\bf r}) = \psi_i({\bf r},t)$ are field operators (either Bose 
or Fermi), $U_i=U_i({\bf r},t)$ is an external potential, $\mu_i$ are chemical 
potentials, and $\Phi_{ij}({\bf r})$ is an interaction potential.

The response functions are the variational derivatives
\be
\label{2}
 \chi_{ij}(\br,t,\br',t') = \frac{\dlt\rho_i(\br,t)}{\dlt U_j(\br',t')}  
\ee
of the density
\be
\label{3}
 \rho_i(\br,t) = \lgl \;  \psi_i^\dgr(\br,t) \; \psi_i(\br,t) \; \rgl
= \pm i G_i(\br,t,\br,t) \; ,
\ee
where we assume the causal Green functions
\be
\label{4}
 G_i(\br,t,\br',t') = - i \; \lgl \; \hat T \psi_i(\br,t) \; 
\psi_i^\dgr(\br',t') \; \rgl \; .
\ee
The Green functions at coinciding times are defined as
\be
\label{5}
 G_i(\br,t,\br',t) = \lim_{t'\ra t+0} \; G_i(\br,t,\br',t') \; .
\ee

Some interaction potentials are known to be non-integrable and not allowing 
for their Fourier transformation. This obstacle can be overpassed by accepting 
an effective interaction potential smoothed by the pair correlation function 
\cite{Kirkwood_31}. It has been shown \cite{Yukalov_7,Yukalov_32,Yukalov_33} 
that, starting from such a correlated approximation, it is possible to formulate 
a self-consistent iterative theory of all orders. Keeping this in mind, we 
assume that the interaction potential is integrable and allows for the Fourier 
transformation
\be
\label{6}
 \Phi_{ij}(\br) = \int \overline \Phi_{ij}(\bk) \; e^{i\bk\cdot\br} \;
\frac{d\bk}{(2\pi)^3} \;  , \qquad
\overline \Phi_{ij}(\bk) = \int \Phi_{ij}(\br) \; e^{-i\bk\cdot\br} \;
d\br \; , 
\ee
with the Fourier transform (\ref{6}) existing for all $k$, including $k=0$. 

Considering a uniform system in volume $V$, we have the constant densities of 
the components
\be
\label{7}
 \rho_j \equiv \frac{N_j}{V} \qquad \left( N = \sum_j N_j \right) \; ,
\ee
hence the constant total density
\be
\label{8}
\rho \equiv \frac{N}{V} = \sum_j \rho_j  \; .
\ee

Assuming that the single-particle spectrum is real, for the Fourier transform 
of the Green function (\ref{4}) we have
\be
\label{9}
G_i(\bk,\om) = \frac{1\pm n_i(\bk)}{\om-\om_i(\bk)+i0} \; \mp \;
\frac{n_i(\bk)}{\om-\om_i(\bk)-i0} \; ,
\ee
with the single-particle spectrum given by the equation
\be
\label{10}
 \om_i(\bk) = \frac{k^2}{2m} + \Sigma_i(\bk,\om_i) - \mu_i \;  ,   
\ee
where $\Sigma_i({\bf k},\omega)$ is a self-energy and the momentum distribution 
is
\be
\label{11}
 n_i(\bk) = \frac{1}{\exp\{\bt \om_i(\bk)\} \mp 1} \qquad 
(\bt T = 1 ) \; ,
\ee
with $T$ being temperature. Here the upper sign is for Bose-Einstein statistics, 
while the lower sign, for Fermi-Dirac statistics. The chemical potentials are 
defined by the normalization to the density
\be
\label{12}
 \rho_i = \int  n_i(\bk) \; \frac{d\bk}{(2\pi)^3} \;  .
\ee

\section{Random-phase approximation}

The response functions are usually calculated in the random-phase approximation 
\cite{Kadanoff_34} with the self-energy taken in the Hartree form
\be
\label{13}
 \Sigma_i(\bk,\om) = \sum_j \Phi_{ij}\; \rho_j \;  ,
\ee
where
\be
\label{14}
 \Phi_{ij} \equiv \int \Phi_{ij}(\br) \; d\br = 
\overline\Phi_{ij}(0) \;  .
\ee

Then the Fourier transforms of the mixture response functions satisfy the 
equations
\be
\label{15}
\chi_{ij}(\bk,\om) = \Pi_{ij}(\bk,\om) + 
\Pi_{ii}(\bk,\om) \sum_n \overline\Phi_{in}(\bk) \; \chi_{nj}(\bk,\om) \;  ,
\ee
with the polarization functions
\be
\label{16}
\Pi_{ij}(\bk,\om) = \pm i \dlt_{ij} \int G_i(\bk+\bk',\om+\om') \;
G_i(\bk',\om')\; \frac{d\bk' d\om'}{(2\pi)^4} \;   .
\ee
For the Green function (\ref{9}), we get
\be
\label{17}
 \Pi_{ii}(\bk,\om) = \frac{k^2}{m_i} \int
\frac{n_i(\bk')}{(\om-\bk\cdot\bk'/m_i)^2-(k^2/2m_i)^2} \;
\frac{d\bk'}{(2\pi)^3} \;  .
\ee

In the case of two components, the solutions to these equations are
$$
\chi_{11} = \frac{\Pi_{11}(1-\Pi_{22}\overline\Phi_{22})}
{(1-\Pi_{11}\overline\Phi_{11})(1-\Pi_{22}\overline\Phi_{22})-
\Pi_{11}\;\Pi_{22}\overline\Phi_{12}^2} \; ,
$$
\be
\label{18}
\chi_{12} = \frac{\Pi_{11}\Pi_{22}\overline\Phi_{12}}
{(1-\Pi_{11}\overline\Phi_{11})(1-\Pi_{22}\overline\Phi_{22})-
\Pi_{11}\;\Pi_{22}\overline\Phi_{12}^2} \; .
\ee
Here, for compactness, the variables ${\bf k}$ and $\omega$ are omitted. The 
response functions $\chi_{21}$ and $\chi_{22}$ can be obtained by exchanging 
the indices $1$ and $2$ in the above expressions. 
   
We consider here an arbitrary potential $\Phi({\bf r})$. In particular, for 
dilute systems of trapped atoms, one has 
$$
\Phi_{ii}= 4\pi \; \frac{a_i}{m_i} \; , \qquad
\Phi_{ij}= 2\pi \; \frac{a_{ij}}{m_{ij}} \;  \qquad 
\left( m_{ij} \equiv \frac{m_i m_j}{m_i + m_j} \right) \;  ,
$$
where $a_i$ and $a_{ij}$ are scattering lengths.

\section{Excitation spectrum}

The spectrum of collective excitations is given by the poles of the response 
functions, that is by the equation
\be
\label{19}
 \chi_{ij}^{-1}(\bk,\om) = 0 \;  .
\ee    
It is convenient to introduce the function
\be
\label{20}
 f_i(\bk,\om) \equiv \frac{1}{\rho_i} \int
\frac{[\om^2-(k^2/2m_i)^2] n_i(\bk')}{(\om-\bk\cdot\bk'/m_i)^2-(k^2/2m_i)^2}\;
\frac{d\bk'}{(2\pi)^3}  
\ee
satisfying the limits
\be
\label{21}
\lim_{k\ra 0} f_i(\bk,\om) = \lim_{k\ra\infty} f_i(\bk,\om) = 1 \;  .
\ee
 
For the mixture of two components, the spectrum of collective excitations can be 
represented \cite{Yukalov_35} as
\be
\label{22}
 \ep^2_\pm = \frac{1}{2} \; \left[ \; \ep_1^2 + \ep_2^2 \; \pm \;
\sqrt{(\ep_1^2-\ep_2^2)^2 + 4\ep_{12}^4} \;\right] \; ,
\ee
where  
\be
\label{23}
 \ep_i^2(\bk) = s_i^2(\bk) \; k^2 + 
\left( \frac{k^2}{2m_i}\right)^2 \;  ,
\ee
with the effective sound velocity
\be
\label{24}
 s_i^2(\bk) = \frac{\rho_i}{m_i} \; 
\overline\Phi_{ii}(\bk) f_i(\bk,\ep_\pm) \;  ,
\ee
and where
\be
\label{25}
\ep_{12}^2(\bk) = s_{12}^2(\bk) \; k^2 \;  ,
\ee
with
\be
\label{26}
s_{12}^2(\bk) =\sqrt{ \frac{\rho_1\rho_2}{m_1 m_2} \; 
f_1(\bk,\ep_\pm)\; f_2(\bk,\ep_\pm) } \; \overline\Phi_{12}(\bk) \; .
\ee

In the long-wave limit, the spectra of a binary mixture reduce to the sound-type 
form
\be
\label{27}
 \ep_\pm(\bk) \simeq s_\pm \; k \qquad ( k \ra 0 ) \; ,
\ee
with the sound velocities given by the equation
\be
\label{28}
s_\pm^2 = \frac{1}{2} \; \left[ \; s_1^2 + s_2^2 \; \pm \;
\sqrt{(s_1^2 - s_2^2)^2 + 4 s_{12}^4 } \; \right ] \;  ,
\ee
where
\be
\label{29}
 s_i^2 \equiv s_i^2(0) = \frac{\rho_i}{m_i} \; \Phi_{ii} \; ,
\qquad 
 s_{12}^2 \equiv s_{12}^2(0) = \sqrt{ \frac{\rho_1\rho_2}{m_1m_2} }\; 
\Phi_{12} \;  .
\ee

From here, it is seen that for the separate stability of each of the components, 
it should be
\be
\label{30}
 \Phi_{ii} > 0 \;  .
\ee
However, the interaction of different components, $\Phi_{12}$ can be of both 
signs. 

In the short-wave limit, one has
\be
\label{31}
\ep_\pm(\bk) \simeq \frac{k^2}{2m_i} \qquad ( k \ra \infty) \; .
\ee

As is clear, the spectrum branch $\varepsilon_+({\bf k})$, describing the 
total density oscillations, is non-negative, and the branch $\ep_-({\bf k})$, 
characterizing the oscillations of the components with respect to each other, 
is non-negative under the condition
\be
\label{32}
 \ep_{12}^2(\bk) < \ep_1(\bk) \; \ep_2(\bk) \;  ,
\ee
which is the condition of mixture stability. In the long-wave limit, we have
\be
\label{33}
 s_{12}^2 < s_1 \; s_2 \;  ,
\ee
which gives the stability condition
\be
\label{34}
 \Phi_{12}^2 < \Phi_{11} \; \Phi_{22} \;  .
\ee
It is interesting that this is the same condition as for a coherent, completely 
Bose-condensed mixture \cite{Yukalov_20}. In general, depending on the 
interaction potentials, there can appear a finite momentum where the stability 
condition (\ref{32}) becomes broken, as under the roton instability in dipolar 
systems \cite{Yukalov_36}.

\section{Moving fluids}

The stability conditions also depend on the motion of fluids 
\cite{Yukalov_20,Yukalov_30,Yukalov_35}. Let us consider a binary mixture, such 
that the components move with velocities ${\bf v}_i$. Then, acting similarly to 
the above analysis, we come to the equation for the spectrum of collective 
excitations
\be
\label{35}
\left[\; ( \om -\al_1 )^2 - \ep_1^2 \; \right] 
\left[\; ( \om -\al_2 )^2 - \ep_2^2 \; \right] = \ep_{12}^4 \; ,
\ee
in which 
\be
\label{36}
 \al_i(\bk) \equiv {\bf v}_i \cdot \bk = v_i \; k \; \cos(\vartheta_i) \; .
\ee
By the Descartes polynomial theorem \cite{Henrici_37} this equation enjoys two 
real roots provided that
\be
\label{37}
 \ep_{12}^4(\bk) <   
\left[\; \ep_1^2(\bk) -\al_1^2(\bk) \; \right] 
\left[\; \ep_2^2(\bk) -\al_2^2(\bk) \; \right]
\ee
for all angles $\vartheta_i$. Minimizing the right-hand side of the above inequality 
yields
\be
\label{38}
\ep_{12}^4(\bk) <   
\left[\; \ep_1^2(\bk) - v_1^2 \; k^2 \; \right] 
\left[\; \ep_2^2(\bk) - v_2^2 \; k^2 \; \right]
\ee
for the angles $\vartheta_i = 0, \pi$. For the broken condition (\ref{38}), under the 
angles $\vartheta_1 = 0$ and $\vartheta_2 = \pi$ or $\vartheta_1 = \pi$ and 
$\vartheta_2 = 0$, this is called the counter-flow instability, and when 
$\vartheta_1 = \vartheta_2$, this is the instability of motion.

In the long-wave limit, inequality (\ref{38}) reduces to the form
\be
\label{39}
 s_{12}^4 < ( s_1^2 - v_1^2 ) ( s_2^2 - v_2^2) \; .
\ee
In terms of the interaction potentials, this is equivalent to the condition
\be
\label{40}
\Phi_{12}^2 < \left( \Phi_{11} \; - \; \frac{m_1 v_1^2}{\rho_1}\right)
\left( \Phi_{22} \; - \; \frac{m_2 v_2^2}{\rho_2}\right) \; .
\ee

\section{Dynamic form-factor}

The measured quantity under neutron or x-ray scattering is the double-differential 
cross section (see \cite{Courteille_5,Yukalov_38})
\be
\label{41}
 \frac{1}{N} \; \frac{d^2\sgm}{d\Om_k\; d\om} = b_{sc}^2 \;
\frac{k_{sc}}{k_{in}} \; S_{tot}(\bk,\om) \;  ,
\ee
in which the transferred momentum and energy are
$$ 
\bk \equiv \bk_{in} - \bk_{sc} \; , \qquad
\om \equiv E_{in} - E_{sc} \; ,
$$
$\Omega_k$ is a solid angle around ${\bf k}$, and $b_{sc}$ is the scattering 
length of a neutron on an atom. The dynamic form-factor at equilibrium is the 
Fourier transform
\be
\label{42}
S_{tot}(\bk,\om) = \frac{1}{N} \sum_{ij} \int
D_{ij}(\br,t,\br',0) \; e^{-i\bk\cdot(\br-\br')+i\om t} \;
d\br d\br' dt  
\ee
of the density-density correlation function
\be
\label{43}
D_{ij}(\br,t,\br',t') \equiv 
\lgl \; \hat n_i(\br,t) \; \hat n_j(\br',t') \; \rgl \; ,
\ee
where the density operator is
\be
\label{44}
\hat n_i(\br,t) \equiv \psi_i^\dgr(\br,t) \; \psi_i(\br,t) \; .
\ee

The total dynamic form-factor is the sum of two terms,
\be
\label{45}
S_{tot}(\bk,\om) = S_{el}(\bk,\om) + S(\bk,\om) \;  .
\ee
The first term
\be
\label{46}
S_{el}(\bk,\om) = \frac{1}{N} \sum_{ij} \int
\rho_i(\br,t) \; \rho_j(\br',0) \; e^{-i\bk\cdot(\br-\br')+i\om t}\;
d\br d\br' dt \; ,
\ee
where
\be
\label{47}
 \rho_i(\br,t) \equiv \lgl \; \hat n_i(\br,t) \; \rgl \; ,
\ee
describes elastic scattering. The second term
\be
\label{48}
S(\bk,\om) = \frac{1}{N} \sum_{ij} \int \left[\;  D_{ij}(\br,t,\br',0) -
\rho_i(\br,t) \; \rho_j(\br',0) \; \right] \; 
e^{-i\bk\cdot(\br-\br')+i\om t}\; d\br d\br' dt \; ,
\ee
also called the van Hove function, describes inelastic scattering.

For a uniform system, where
$$
 \rho_i(\br,t) = \rho_i = \frac{N_i}{V} \;  ,
$$
the elastic scattering term is
\be
\label{49}
 S_{el}(\bk,\om) = ( 2\pi)^4 \sum_{ij} \frac{\rho_i\rho_j}{\rho} \;
\dlt(\bk) \; \dlt(\om) \;  .
\ee
The inelastic term takes the form
\be
\label{50}
 S(\bk,\om) = \sum_{ij} S_{ij}(\bk,\om) \;  ,
\ee
where
\be
\label{51}
S_{ij}(\bk,\om) = \frac{1}{N} \int \left[\;  D_{ij}(\br,t,\br',0) -
\rho_i(\br,t) \; \rho_j(\br',0) \; \right] \; 
e^{-i\bk\cdot(\br-\br')+i\om t}\; d\br d\br' dt \; .
\ee

Defining the Fourier transform of the density-density correlation function
\be
\label{52}
D_{ij}(\bk,\om) = \int D_{ij}(\br,t,0,0)\; 
e^{-i\bk\cdot\br+i\om t}\; d\br dt 
\ee
gives for a uniform system
\be
\label{53}
 S_{ij}(\bk,\om) = \frac{1}{\rho} \; \left[\;  D_{ij}(\bk,\om) -
(2\pi)^4 \; \rho_i \; \rho_j \; \dlt(\bk) \; \dlt(\om) \; \right] \; .
\ee

The density-density correlation function (\ref{43}) is connected with the 
density-density Green function
$$
  - i \; \lgl \; \hat T \; \hat n(\br,t) \; \hat n(\br,t') \; \rgl \;  .
$$
Using the properties of this Green function \cite{Yukalov_39} results in the 
Fourier transform
\be
\label{54}
D_{ij}(\bk,\om) = - \; 
\frac{2{\rm Im}\chi_{ij}(\bk,\om)}{1+ e^{-\bt\om} } +
(2\pi)^4 \; \rho_i \; \rho_j \; \dlt(\bk) \; \dlt(\om) \; ,
\ee
where $\chi_{ij}({\bf k}, \omega)$ is the Fourier transform of the response 
function (\ref{2}). Respectively, for the dynamic form-factor (\ref{51}), we get
\be
\label{55}
 S_{ij}(\bk,\om) = - \; 
\frac{2{\rm Im}\chi_{ij}(\bk,\om)}{\rho(1+ e^{-\bt\om}) } \; .
\ee

The response function enjoys the spectral representation
\be
\label{56}
 \chi_{ij}(\bk,\om) = \int_{-\infty}^\infty 
\frac{\gm_{ij}(\bk,\om')}{\om-\om'} \; \frac{d\om'}{2\pi} \; ,
\ee
with the spectral function
\be
\label{57}
 \gm_{ij}(\bk,\om) = i \; [\; \chi_{ij}(\bk,\om+i0) -
\chi_{ij}(\bk,\om-i0) \; ] \; ,
\ee
and the dispersion relation
\be
\label{58}
 {\rm Im}\chi_{ij}(\bk,\om) = - \; \frac{1}{2} \; \gm_{ij}(\bk,\om) \;
\coth\left( \frac{\bt\om}{2}\right) \; .
\ee
This gives
\be
\label{59}
 S_{ij}(\bk,\om) = \frac{\gm_{ij}(\bk,\om)}{\rho(1-e^{-\bt\om})}\;  .
\ee

The dynamic form-factor possesses the properties
\be
\label{60}
 S_{ij}(-\bk,\om) = S_{ij}(\bk,\om) \; , \qquad 
 S_{ij}(\bk,-\om) = S_{ij}(\bk,\om) \; e^{-\bt\om} \; .
\ee
It can be written as the sum
\be
\label{61}
  S_{ij}(\bk,\om) =  S_{ij}^+(\bk,\om)  +  S_{ij}^-(\bk,-\om)
\ee
of the symmetric term
\be
\label{62}
S_{ij}^+(\bk,\om) \equiv \frac{1}{2} \; 
[\; S_{ij}(\bk,\om) + S_{ij}(\bk,-\om) \; ] = \frac{1}{2\rho} \;
\gm_{ij}(\bk,\om) \; \coth\left( \frac{\bt\om}{2}\right) = 
S_{ij}^+(\bk,-\om)
\ee
and the antisymmetric term
\be
\label{63}
S_{ij}^-(\bk,\om) \equiv \frac{1}{2} \; 
[\; S_{ij}(\bk,\om) - S_{ij}(\bk,-\om) \; ] = \frac{1}{2\rho} \;
\gm_{ij}(\bk,\om)  = - S_{ij}^-(\bk,-\om)
\ee
with respect to the change of the sign of $\omega$.

\section{Response functions}

The response functions (\ref{18}) can be written in the form
$$
 \chi_{11}(\bk,\om) = \left( \frac{\rho_1}{m_1}\right)\;
\frac{[\om^2-\ep_2^2(\bk)] k^2}{[\om^2-\ep_+^2(\bk)][\om^2-\ep_-^2(\bk)]} \; ,
$$
\be
\label{64}
 \chi_{12}(\bk,\om) = \sqrt{\frac{\rho_1\rho_2}{m_1m_2} } \;
\frac{\ep_{12}^2(\bk) k^2}{[\om^2-\ep_+^2(\bk)][\om^2-\ep_-^2(\bk)]} \;  .
\ee
Then the spectral functions (\ref{57}) become
$$
\gm_{11}(\bk,\om) =  \frac{\rho_1}{m_1}\;
[\; \om^2-\ep_2^2(\bk) \; ]\; k^2 \; \gm(\bk,\om) \; ,
$$
\be
\label{65}
\gm_{12}(\bk,\om) =  \sqrt{ \frac{\rho_1\rho_2}{m_1m_2} } \;
\ep_{12}^2(\bk)\; k^2 \; \gm(\bk,\om)  \; ,
\ee
where
$$
\gm(\bk,\om) = 2\pi\dlt \left( \; [ \; \om^2 - \ep_+^2(\bk) \; ]
\; [ \; \om^2 - \ep_-^2(\bk) \; ] \; \right) =
$$
$$
= 
\frac{\pi}{\ep_+^2(\bk) - \ep_-^2(\bk)} \; \left\{
\frac{1}{\ep_+(\bk)} \; [ \; \dlt(\om - \ep_+(\bk) ) - 
\dlt(\om + \ep_+(\bk) ) \; ] \; - \right.
$$
\be
\label{66}
 - \; \left.
\frac{1}{\ep_-(\bk)} \; [ \; \dlt(\om - \ep_-(\bk) ) - 
\dlt(\om + \ep_-(\bk) ) \; ] \; \right\} \; .
\ee

\section{Structure factor}

The system structure factor
\be
\label{67}
S(\bk) = \sum_{ij} S_{ij}(\bk)
\ee
is the sum of the partial structure factors
\be
\label{68}
S_{ij}(\bk) = \int_{-\infty}^\infty S_{ij}(\bk,\om) \;
\frac{d\om}{2\pi} \; .
\ee
For the latter, we have
\be
\label{69}
 S_{ij}(\bk) = \frac{1}{\rho} \int_0^\infty \gm_{ij}(\bk,\om) 
\coth \left( \frac{\bt\om}{2}\right) \; d\om \;  .
\ee
Explicitly, for a binary mixture, this reads as
$$
S_{11}(\bk) = \left( \frac{\rho_1}{m_1}\right) \;
\frac{k^2}{2\rho[\;\ep_+^2(\bk)-\ep_-^2(\bk)\;]} \; \times
$$
\be
\label{70}
\times
\left[ \; \frac{\ep_+^2(\bk)-\ep_2^2(\bk)}{\ep_+(\bk)} \;
\coth\left( \frac{\bt\ep_+(\bk)}{2}\right) \; - \;
\frac{\ep_-^2(\bk)-\ep_2^2(\bk)}{\ep_-(\bk)} \;
\coth\left( \frac{\bt\ep_-(\bk)}{2}\right) \; \right]
\ee
and
$$
S_{12}(\bk) = \sqrt{ \frac{\rho_1\rho_2}{m_1 m_2} } \;
\frac{\ep_{12}^2(\bk) k^2}{2\rho[\ep_+^2(\bk)-\ep_-^2(\bk)]} \;
\times
$$
\be
\label{71}
 \times
\left[\; 
\frac{1}{\ep_+(\bk)}\; \coth\left( \frac{\bt\ep_+(\bk)}{2} \right) 
\; - \;
\frac{1}{\ep_-(\bk)}\; \coth\left( \frac{\bt\ep_-(\bk)}{2} \right) 
\; \right] \; .
\ee

It is possible to check that in the short-wave limit
\be
\label{72}
 \lim_{k\ra\infty} \;\sum_{ij} S_{ij}(\bk) = 1 \; .
\ee

The limiting behavior of the partial structure factors depends on the 
relation between the sound modes and temperature. If the sound-mode energy 
is much smaller than temperature, then the structure factors tend to the 
limits
$$
S_{11}(0) = \left( \frac{\rho_1}{m_1}\right) \; 
\frac{s_2^2(0) T}{\rho s_+^2(0) s_-^2(0)} \; , 
$$
\be
\label{73}
S_{12}(0) = \sqrt{ \frac{\rho_1\rho_2}{m_1 m_2} } \;
\frac{s_{12}^2(0) T}{\rho s_+^2(0) s_-^2(0)} 
\qquad
\left( \frac{s_\pm k}{T} \ll 1 \right) \; .
\ee
In the opposite limit of temperature much smaller than the sound-mode energy, 
one has
$$
S_{11}(k) \simeq \left( \frac{\rho_1}{m_1}\right) \;
\frac{[\ep_+(\bk)\ep_-(\bk)+\ep_2^2(\bk)]k^2}
{2\rho\ep_+(\bk)\ep_-(\bk)[\ep_+(\bk)+\ep_-(\bk)]} \; ,
$$
\be
\label{74}
S_{12}(k) \simeq -\; \sqrt{ \frac{\rho_1\rho_2}{m_1 m_2} } \;
\frac{\ep_{12}^2(\bk) k^2}
{2\rho\ep_+(\bk)\ep_-(\bk)[\ep_+(\bk)-\ep_-(\bk)]} \;  ,
\ee
provided that
$$
 \frac{T}{s_\pm k} \ll 1 \;  .
$$
If both temperature and the sound-mode energy tend to zero so that $T/s_{\pm}k$ 
tends to zero, then
$$
 S_{11}(\bk) \simeq   \left( \frac{\rho_1}{m_1}\right) \;
\frac{[s_+(0) s_-(0)+s_2^2(0)] k}
{2\rho s_+(0) s_-(0)[s_+(0)+s_-(0)]} \; ,
$$
\be
\label{75}
S_{12}(\bk) \simeq -\;\sqrt{ \frac{\rho_1\rho_2}{m_1 m_2} } \;
\frac{s_{12}^2(0) k}{2\rho s_+(0) s_-(0)[s_+(0)+s_-(0)]} \; ,
\ee
under the condition
$$
 T \ra 0 \; , \qquad k \ra 0 \; , 
\qquad
\frac{T}{s_\pm k} \ra 0 \;  .
$$
In any case, when the mixture looses stability and starts separating, with 
$s_-(0)$ tending to zero, the structure factors diverge.

\section{Sum rules}

One considers the integral
\be
\label{76}
K_{ij}(\bk) = \int_{-\infty}^\infty \om S_{ij}(\bk,\om) \; 
\frac{d\om}{2\pi}
\ee
that, using the properties of the dynamic form-factor, can be written as
\be
\label{77}
 K_{ij}(\bk) = \frac{1}{\rho} \int_0^\infty \om \gm_{ij}(\bk,\om) \; 
\frac{d\om}{2\pi} \; .
\ee
With the spectral functions (\ref{65}), this yields
\be
\label{78}
 K_{ij}(\bk) = \dlt_{ij} \; \frac{\rho_i}{\rho} \;
\left( \frac{k^2}{2m_i} \right) \;  .
\ee

The other important integral is
\be
\label{79}
 Q_{ij}(\bk) = \int_{-\infty}^\infty \frac{1}{\om}\; S_{ij}(\bk,\om) \; 
\frac{d\om}{2\pi} \;   ,
\ee
which can be rewritten as 
\be
\label{80}
 Q_{ij}(\bk) = \frac{1}{\rho} \int_0^\infty \frac{1}{\om} \; 
\gm_{ij}(\bk,\om) \; \frac{d\om}{2\pi}    .
\ee
Taking into account the spectral representation (\ref{56}) gives
\be
\label{81}
 Q_{ij}(\bk) = -\; \frac{1}{\rho} \; {\rm Re}\;\chi_{ij}(\bk,0) \; .
\ee

Since the response function has the spectral representation
\be
\label{82}
\chi_{ij}(\bk,\om) = \int_{-\infty}^\infty \gm_{ij}(\bk,\om') \;
\left[\; \frac{1+n(\om')}{\om-\om'+i0} \; - \;
\frac{n(\om')}{\om-\om'-i0} \; \right] \; \frac{d\om'}{2\pi} \;  , 
\ee
where $n(\omega) \equiv (e^{\beta \omega}-1)^{-1}$, then using the property
$$
 \frac{1}{\om\pm i0} = P \; \frac{1}{\om} \; \mp \; i \pi \dlt(\om) \; , 
$$
where $P$ implies the principal value, we obtain the real part
\be
\label{83}
 {\rm Re} \; \chi_{ij}(\bk,\om) = P  \int_{-\infty}^\infty
\frac{\gm_{ij}(\bk,\om')}{\om-\om'} \; \frac{d\om'}{2\pi}
\ee
and the imaginary part
\be
\label{84}
 {\rm Im} \; \chi_{ij}(\bk,\om) = -\; \frac{1}{2} \; \gm_{ij}\bk,\om) \;
[\; 1 + 2 n(\om) \; ] \;  .
\ee
Here
$$
1 + 2 n(\om) = \frac{e^{\bt\om}+1}{e^{\bt\om}-1} = 
\coth\left( \frac{\bt\om}{2} \right) \; .
$$

\section{Isothermal compressibility}

The response functions define the compressibility functions
\be
\label{85}
\varkappa_i(\br,t,\br',t') \equiv -\; 
\frac{\chi_{ii}(\br,t,\br',t')}{\rho_i(\br,t)\rho_i(\br',t') } \; ,
\ee
whose Fourier transforms are
\be
\label{86}
\varkappa_i(\bk,\om) = 
\int \chi_i(\br,t,0,0) \; e^{-i\bk\cdot\br+i\om t} \; d\br dt \; .
\ee
For an equilibrium uniform system, we have
\be
\label{87}
\varkappa_i(\bk,\om) = - \; \frac{\chi_{ii}(\bk,\om)}{\rho_i^2} \; .
\ee

The coefficient of isothermal compressibility is defined by the expression 
\be
\label{88}
 \varkappa_i \equiv {\rm Re}\; \varkappa_i(0,0) = - \;
\frac{{\rm Re} \chi_{ii}(0,0)}{\rho_i^2} \;  .
\ee
Hence it is given by the relation
\be
\label{89}
\varkappa_i = \frac{\rho}{\rho_i^2} \; Q_{ii}(0) \;  .
\ee

For a binary mixture, we get
$$
{\rm Re}\; \chi_{11}(0,0) = - \; 
\frac{\rho_1 s_2^2(0)}{m_1 s_+^2(0) s_-^2(0)} \; ,
$$
\be
\label{90}
{\rm Re}\; \chi_{12}(0,0) = \sqrt{ \frac{\rho_1\rho_2}{m_1 m_2} } \;
\frac{s_{12}^2(0)}{s_+^2(0) s_-^2(0)} \;  .
\ee
Therefore
\be
\label{91}
 \varkappa_1 = \frac{s_2^2(0)}{\rho_1 m_1 s_+^2(0) s_-^2(0)} \; ,
\qquad
\varkappa_2 = \frac{s_1^2(0)}{\rho_2 m_2 s_+^2(0) s_-^2(0)} \; .
\ee
Thus we find another relation for the isothermal compressibilities
\be
\label{92}
 \varkappa_i = \frac{\rho S_{ii}(0)}{\rho_i^2 T} \;  .
\ee

When the mixture looses its stability and $s_-(0)$ tends to zero, the 
compressibility coefficients diverge.

\section{Thermodynamic stability}

Thermodynamic stability conditions are based on the comparison of the free energies 
of the mixed, $F_{mix}$, and separated, $F_{sep}$, systems. The mixture is stable 
when
\be
\label{93}
 F_{mix} < F_{sep} \; .
\ee
In what follows, we consider a uniform system with the component densities 
$\rho_i({\bf r}) = N_i/V$ and use the Hartree approximation.  

The free energy of a mixed system writes as
\be
\label{94}
  F_{mix} = E_{mix} - T S_{mix} \;  .
\ee
The energy 
\be
\label{95}
E_{mix} = \lgl \; H \; \rgl + \sum_i \mu_i N_i =
K_{mix} + E_{mix}^{ext} + E_{mix}^{int}
\ee
includes the kinetic energy term $K_{mix}$, the energy due to external 
potentials,
\be
\label{96}
E_{mix}^{ext} = \sum_i \int_V U_i(\br) \; \rho_i(\br) \; d\br =
\sum_i U_i \; N_i \; ,
\ee
where
$$
 U_i \equiv \frac{1}{V} \int_V U_i(\br) \; d\br \;  ,
$$
and the interaction energy 
\be
\label{97}
 E_{mix}^{int} = \frac{1}{2} \sum_{ij} \int_V  
\rho_i(\br) \; \Phi_{ij}(\br-\br') \; \rho_j(\br') \; d\br d\br' \; .
\ee
For a uniform system, the latter is
\be
\label{98}
 E_{mix}^{int} = \frac{1}{2} 
\sum_{ij} \Phi_{ij} \; \frac{N_i N_j}{V} \; ,
\ee
where
\be
\label{99}
 \Phi_{ij} \equiv \int_V \Phi_{ij}(\br) \; d\br \; .
\ee

The Hamiltonian of a separated system is
$$
 H_{sep} = \sum_i \int \psi_i^\dgr(\br) \; \left( - \;
\frac{\nabla^2}{2m} + U_i - \mu_i\right) \psi_i(\br) \; d\br \; + 
$$
\be 
\label{100}
+ \;
\frac{1}{2} \sum_i \int \psi_i^\dgr(\br) \; \psi_i^\dgr(\br') \;
\Phi_{ii}(\br-\br') \; \psi_i(\br') \; \psi_i(\br) \; d\br d\br' \; .
\ee
The free energy writes as
\be
\label{101}
F_{sep} = E_{sep} - T S_{sep} \;   .
\ee
The energy
\be
\label{102}
E_{sep} = \lgl \; H_{sep} \; \rgl + \sum_i \mu_i N_i =
K_{sep} + E_{sep}^{ext} + E_{sep}^{int}
\ee
contains the kinetic energy $K_{sep}$, the energy caused by external potentials 
$E_{sep}^{ext}$, and the interaction energy $E_{sep}^{int}$. 

The assumption of the system uniformity presupposes that the system volumes are 
sufficiently large, so that the equalities are valid:
$$
\frac{1}{V} \int_V U_i(\br)\; d\br = 
\frac{1}{V_i} \int_{V_i} U_i(\br)\; d\br \; ,
\qquad
\int_V \Phi_{ii}(\br)\; d\br = \int_{V_i} \Phi_{ii}(\br)\; d\br \; .
$$
Also, we assume that the difference in the kinetic energy, under the same given 
conditions, can be neglected as compared to the difference in the potential 
energy, thus setting $K_{mix}-K_{sep} \approx 0$. The component densities of 
the uniform separated system are
\be
\label{103}
 \rho_i^{loc}(\br) = \lgl \; \psi_i^\dgr(\br)\; \psi_i(\br)\; \rgl =
\frac{N_i}{V_i} \;  .
\ee

Then the energy due to external potentials is
\be
\label{104}
E_{sep}^{ext} = \sum_i \int_{V_i} U_i(\br)\; \rho_i^{loc}(\br)\; d\br =
\sum_i U_i \; N_i \;  ,
\ee
while the interaction energy of the separated system becomes
\be
\label{105}
E_{sep}^{int} = \frac{1}{2} \sum_i \int_{V_i} \rho_i^{loc}(\br)\; 
\Phi_{ii}(\br-\br') \; \rho_i^{loc}(\br')\; d\br d\br' \; ,
\ee
which gives
\be
\label{106}
E_{sep}^{int} = \frac{1}{2} \sum_i  \Phi_{ii} \; \frac{N_i^2}{V_i} \; .
\ee

The condition of thermodynamic stability (\ref{93}) takes the form
\be
\label{107}
 E_{mix}^{int} - E_{sep}^{int} < N\; T\; \Dlt S_{mix} \; ,
\ee
in which
\be
\label{108}
 \Dlt S_{mix} = \frac{1}{N} \; ( S_{mix} - S_{sep} )  
\ee
is the entropy of mixing per particle.   

Mechanical equilibrium implies the equality of the pressures
\be
\label{109}
P_i = - \; \frac{\prt F_{sep}}{\prt V_i} = P_j = - \; 
\frac{\prt F_{sep}}{\prt V_j}   .
\ee
From the latter condition, we have
$$
 \frac{V_j}{V_i} = 
\frac{N_j}{N_i} \; \sqrt{ \frac{ \Phi_{jj}}{\Phi_{ii}} } \;  .
$$
This allows us to rewrite the stability condition (\ref{107}) as the inequality
\be
\label{110}
 \sum_{ij} \frac{N_i N_j}{2V} \;
\left( \Phi_{ij} \; - \; \sqrt{\Phi_{ii} \Phi_{jj} } \; - \;
\frac{2T}{\rho} \; \Dlt S_{mix} \right) < 0 \; .
\ee
Assuming that this inequality is valid for any $N_i$ and $N_j$, we come to the
stability condition \cite{Yukalov_30}
\be
\label{111}
 \Phi_{ij} \; - \; \sqrt{\Phi_{ii}\; \Phi_{jj} } \; < \;
\frac{2T}{\rho} \; \Dlt S_{mix} \;  .
\ee
The entropy of mixing per particle can be represented as
\be
\label{112}
 \Dlt S_{mix} = - \sum_i n_i \; \ln n_i \qquad
\left( n_i \equiv \frac{N_i}{N} \right) \; .
\ee
 
The mixture stability condition (\ref{111}) shows that finite temperatures facilitate
the component mixing. The components could be immiscible at zero temperature but become 
miscible at finite temperature. This is also in agreement with the miscibility criterion 
for Bose-condensed mixtures, taking into account strong interactions and temperature 
\cite{Rakhimov_26}. The miscibility condition of type (\ref{34}) can be invalid, but 
the components, anyway, can be miscible at finite temperatures and interactions.

\section{Conclusion}

Mixtures of quantum fluids, liquids or gases, are considered. The studied systems 
are treated as normal quantum systems, that is requiring quantum techniques for their
description, but exhibiting no superfluidity or superconductivity. The main point of
interest in the present review is the conditions of quantum mixture stability with 
respect to the spatial separation into different components. The mixture stability 
can be studied from the point of view of dynamic or thermodynamic stability.   

The study of dynamic stability is based on the investigation of spectra of collective 
excitations. The principal difference between Bose-condensed and normal fluids is in 
the following. In a Bose-condensed system, the global gauge symmetry is broken. This 
symmetry breaking is conveniently characterized by the Bogolubov shift 
\cite{Bogolubov_40,Bogolubov_41,Bogolubov_42}. Then the vacuum state is described by 
a coherent state formed by completely condensed particles, while the uncondensed 
particles play the role of excitations above the vacuum \cite{Yukalov_43}. Because 
of this, the single-particle Green functions of uncondensed particles, actually, 
describe excitations above the coherent-state vacuum. This is why to study the 
collective spectrum of a Bose-condensed system it is sufficient to consider solely 
the single-particle Green functions of uncondensed particles, whose poles give the 
required collective spectrum. 

On the other hand, the vacuum for a normal quantum system is an empty state. The poles
of single-particle Green functions describe the single-particle spectra, but the 
spectrum of collective excitations is characterized by the poles of two-particle Green
functions or by the poles of response functions.

For the description of response functions, one usually involves the random-phase
approximation. This approximation describes well the low-temperature states, but the
influence of temperature on the spectrum of collective excitations is not as important. 
 
The role of temperature can be investigated by resorting to the study of thermodynamic 
stability, when one compares the free energies of a mixed and separated systems. The
found stability criterion demonstrates that temperature facilitates the mixing of
different quantum components. At low temperature, the system could be immiscible, 
while becomes miscible when rising temperature. 

This effect also is in agreement with the study of stability of Bose-condensed systems 
with strong interactions and finite temperature. For the correct description of such 
systems, it is necessary to take into account anomalous averages and the existence of
two chemical potentials for each of the components, which is caused by the global gauge 
symmetry breaking \cite{Yukalov_44,Yukalov_45,Yukalov_46}. Including into the description 
the corrections above the coherent approximation implies the inclusion into consideration 
of quantum and thermal fluctuations. These fluctuations make the mixing easier. The 
Bose-condensed mixture stability, accurately taking account of fluctuations, has been 
recently analyzed \cite{Rakhimov_26}, confirming that fluctuations facilitate the mixing 
of the components.


\newpage


\begin{thebibliography}{99}


\bibitem{Letokhov_1}
Letokhov V 2007 
{\it Laser Control of Atoms and Molecules} 
(New York: Oxford University)

\bibitem{Lieb_2}
Lieb E H, Seiringer R, Solovej J P and Yngvason J 2005 
{\it The Mathematics of the Bose Gas and its Condensation} 
(Basel: Birkhauser)

\bibitem{Pethick_3}
Pethick C J and Smith H 2008 
{\it Bose-Einstein Condensation in Dilute Gases} 
(Cambridge: Cambridge University)

\bibitem{Ueda_4}
Ueda M 2010 
{\it Fundamentals and New Frontiers of Bose-Einstein Condensation} 
(Singapore: World Scientific)

\bibitem{Courteille_5}
Courteille P W, Bagnato V S and Yukalov V I 2001 
Laser Phys. {\bf 11} 659

\bibitem{Andersen_6}
Andersen J O 2004 
{\it Rev. Mod. Phys.} {\bf 76} 599

\bibitem{Yukalov_7}
Yukalov V I 2004 
{\it Laser Phys. Lett.} {\bf 1} 435

\bibitem{Bongs_8}
Bongs K and Sengstock K 2004 
{\it Rep. Prog. Phys.} {\bf 67} 907

\bibitem{Yukalov_9}
Yukalov V I and Girardeau M D 2005 
{\it Laser Phys. Lett.} {\bf 2} 375

\bibitem{Posazhennikova_10}
Posazhennikova A 2006 
{\it Rev. Mod. Phys.} {\bf 78} 1111

\bibitem{Morsch_11}
Morsch O and Oberthaler M 2006 
{\it Rev. Mod. Phys.} {\bf 78} 179

\bibitem{Yukalov_12}
Yukalov V I 2007 
{\it Laser Phys. Lett.} {\bf 4} 632

\bibitem{Moseley_13}
Moseley C, Fialko O and Ziegler K 2008 
{\it Ann. Phys. (Berlin)} {\bf 17} 561

\bibitem{Bloch_14}
Bloch I, Dalibard J and Zwerger W 2008 
{\it Rev. Mod. Phys.} {\bf 80} 885

\bibitem{Proukakis_15}
Proukakis N P and Jackson B 2008 
{\it J. Phys. B} {\bf 41} 203002

\bibitem{Yurovsky_16}
Yurovsky V A, Olshanii M and Weiss D S 2008 
{\it Adv. At. Mol. Opt. Phys.} {\bf 55} 61

\bibitem{Castilho_17}
Castilho P C M, E Pedrozo-Penafiel E,  Gutierrez E M, Mazo P L, Roati G, 
Farias K M and Bagnato V S 2019
{\it Laser Phys. Lett.} {\bf 16} 035501 

\bibitem{Saarela_18}
Saarela M and Taipaleenmaki T 2003
{\it Int. J. Mod. Phys. B} {\bf 17} 5227

\bibitem{Adhikari_19}
Adhikari S K 2007
{\it Phys. Rev. A} {\bf 76} 053609

\bibitem{Yukalov_20}
Yukalov V I and Yukalova E P 2004
{\it Laser Phys. Lett.} {\bf 1} 50

\bibitem{Abad_21}
Abad M, Recati1 A, Stringari1 S and Chevy F 2015
{\it Eur. Phys. J. D} {\bf 69} 126

\bibitem{Nepomnyashchii_22}
Nepomnyashchii Y A 1974
{\it Theor. Math. Phys.} {\bf 20} 399

\bibitem{Hugenholtz_23}
Hugenholtz N M and Pines D 1959
{\it Phys. Rev.} {\bf 116} 489

\bibitem{Nepomnyashchii_24}
Nepomnyashchii Y A 1976
{\it J. Exp. Theor. Phys.} {\bf 43} 559

\bibitem{Watabe_25}
Watabe S 2021
{Phys. Rev. A} {\bf 103} 053307

\bibitem{Rakhimov_26}
Rakhimov A, Abdurakhmonov T, Narzikulov Z and Yukalov V I 2022
{\it Phys. Rev. A} {\bf 106} 033301

\bibitem{Gavoret_27}
Gavoret T and Nozi\`{e}res P 1964
{\it Ann. Phys. (N.Y.)} {\bf 28} 349

\bibitem{Watabe_28}
Watabe S 2020
{\it New J. Phys.} {\bf 22} 103010

\bibitem{Kurten_29}
K\"{u}rten K E and Campbell C E 1982
{\it Phys. Rev. B} {\bf 26} 124

\bibitem{Yukalov_30}
Yukalov V I 2016
{\it Laser Phys.} {\bf 26} 062001

\bibitem{Kirkwood_31}
Kirkwood J G 1965
{\it Quantum Statistics and Cooperative Phenomena}
(New York: Gordon and Breach)

\bibitem{Yukalov_32}
Yukalov V I 1990
{\it Phys. Rev. A} {\bf 42} 3324

\bibitem{Yukalov_33}
Yukalov V I 2016
{\it Phys. Rev. E} {\bf 94} 102106

\bibitem{Kadanoff_34}
Kadanoff L P and Baym G 1962
{\it Quantum Statistical Mechanics} (New York: Benjamin)

\bibitem{Yukalov_35}
Yukalov V I 1980
{\it Acta Phys. Pol. A} {\bf 57} 295

\bibitem{Yukalov_36}
Yukalov V I 2018
{\it Laser Phys.} {\bf 28} 053001

\bibitem{Henrici_37}
Henrici P 1988 
{\it Applied and Computational Complex Analysis} Vol. 1
(New York: Wiley) 

\bibitem{Yukalov_38}
Yukalov V I 2007
{\it J. Phys. Stud.} {\bf 11} 55

\bibitem{Yukalov_39}
Yukalov V I 1998
{\it Statistical Green's Functions} 
(Kingston: Queen's University)

\bibitem{Bogolubov_40}
Bogolubov N N 1967
{\it Lectures on Quantum Statistics} 
(New York: Gordon and Breach) Vol. 1

\bibitem{Bogolubov_41}
Bogolubov N N 1970 
{\it Lectures on Quantum Statistics} 
(New York: Gordon and Breach) Vol. 2

\bibitem{Bogolubov_42}
Bogolubov N N 2015
{\it Quantum Statistical Mechanics} 
(Singapore: World Scientific)

\bibitem{Yukalov_43}
Yukalov V I 2006
{\it Laser Phys.} {\bf 16} 511

\bibitem{Yukalov_44}
Yukalov V I 2005
{\it Phys. Rev. E} {\bf 72} 066119

\bibitem{Yukalov_45}
Yukalov V I 2006
{\it Phys. Lett. A} {\bf 359} 712

\bibitem{Yukalov_46}
Yukalov V I 2008
{\it Ann. Phys. (N.Y.)} {\bf 323} 461

\end{thebibliography}
\end{document}